# Entanglement, and unsorted database search in noise-based logic

**Laszlo Kish [1,*] and Walter C. Daugherity [2]**

[1] Department of Electrical and Computer Engineering, Texas A&M University, 3128 TAMU, College Station, TX 77843-3128; Laszlokish@tamu.edu
[2] Department of Computer Science and Engineering, Texas A&M University, 3112 TAMU, College Station, TX 77843-3112; daugher@tamu.edu
* Correspondence: Laszlokish@tamu.edu

**Featured Application: Exponential speedup with special-purpose computational tasks, similar to quantum computing claims. Though the goals are similar to those of quantum computing, the way of realization is substantially different and free of some of the negative effects such as decoherence and errors. Most of the particular solutions described here are free from the statistical nature and its related slowdown of quantum computing. This fact supports the opinion that the claimed advantages of quantum computers come from the stochastic nature of quantum measurements. It is well-known in the field of random algorithms that stochasticity can offer exponential speedup for certain special-purpose operations. In the same way, classical stochasticity via the instantaneous noise-based logic schemes may offer more than quantum computers because of the lack of the fundamental disadvantages of quantum systems which are the decoherence, the statistical nature of quantum measurements, and the fact that measurements destroy the state. Examples of this superiority are shown in the present paper where, once the search has begun, the results come out deterministically without statistical errors, resulting in exponential speedup compared to Grover's quantum search algorithm.**

**Abstract:** We explore the collapse of "wavefunction" and the measurement of entanglement in the superpositions of hyperspace vectors in classical physical instantaneous-noise-based logic (INBL). We find both similarities with and major differences from the related properties of quantum systems. Two search algorithms utilizing the observed features are introduced. For the first one we assume an unsorted names database set up by Alice that is a superposition (unknown by Bob) of up to $n$=2^$N$ strings; those we call names. Bob has access to the superposition wave and to the 2$N$ reference noises of the INBL system of $N$ noise bits. For Bob, to decide if a given name $x$ is included in the superposition, once the search has begun, it takes $N$ switching operations followed by a single measurement of the superposition wave. Thus the time and hardware complexity of the search algorithm is O[log($n$)] which indicates an exponential speedup compared to Grover's quantum algorithm in a corresponding setting. An extra advantage is that the error probability of the search is zero. Moreover, the scheme can also check the existence of a fraction of a string, or several separate string fractions embedded in an arbitrarily long, arbitrary string. In the second algorithm, we expand the above scheme to a phonebook with $n$ names and $s$ phone numbers. When the names and numbers have the same bit resolution, once the search has begun, the time and hardware complexity of this search algorithm is O[log($n$)]. In the case of one-to-one correspondence between names and phone numbers ($n$=$s$), the algorithm offers inverse phonebook search too. The error probability of this search algorithm is also zero.

## 1. Introduction: On noise-based logic

The Einstein-Podolsky-Rosen paradox [1] points to a deep feature of quantum physics leading to the notion of entanglement. In this paper, we address a similar problem in noise-based logic (NBL) [2-15], which is a classical physical scheme where noise-bits carry the same high-dimensional Hilbert space as qubits in quantum computing. We explore some of the similarities and differences between the entanglements and their measurements in these systems and propose relevant data search applications for NBL.

Noise-based logic (NBL) [2-15] is a classical physical computation system using orthogonal stochastic processes to carry and process the logic information. Even though the creation of NBL was originally inspired by the stochastic nature of neural signals [2-5] and there are some brain-inspired NBL schemes [6,7] that may help in understanding the functioning of the biological brain, the developed NBL schemes and their methods reach far beyond the brain-mimicking efforts [8-15]. Similarly, there is some resemblance between NBL and quantum computing including some exponentially large superpositions that can be created by polynomial complexity [8-15]. However the NBL schemes and their modes of operations are substantially different, even if some of the computational goals may be the same. While the explorations of NBL are still taking place, it is already clear that in certain demonstrated cases the robustness and controllability of classical physical systems and information offer strong advantages [14,15] over quantum informatics. Some NBL schemes can offer solutions that either reach beyond the abilities of quantum informatics or offer alternative means with identical potential and less cost, similarly to recent achievements by Ewin Tang in classical computing [16,17]. The present paper is an additional new example of this fact.

Note that the exploration of classical physical stochastic processes for information technology also yielded very similar positive surprises regarding quantum encryption. The classical statistical physical key exchange scheme, the KLJN system [18], offers the same level of security and it is the only currently known unconditionally secure communication scheme that can be integrated on a chip.

### *1.1. Instantaneous noise-based logic*

In this paper we explore instantaneous noise-based logic (INBL) [4,5,8-15] for search algorithms over unknown superpositions within the NBL framework.

A system of $M$ noise-bits consists of $2M$ uncorrelated, statistically independent, true random noise generators, with zero mean – the *reference noise system* – to supply stochastic signals (*reference noises*),

$$R_{10}(t), R_{11}(t), R_{20}(t), R_{21}(t), ..., R_{M0}(t), R_{M1}(t) \quad , \qquad (1)$$

that represent the Low (0) and High (1) values of each noise-bit. These signals are uncorrelated and have zero mean value. The NBL gates and processors utilize these reference noises together with the input data for algebraic operations to produce their output. Therefore the reference noises are available for each unit in the NBL processor and *reference wires* indicate their supply in the circuit drawings, see Fig. 1.

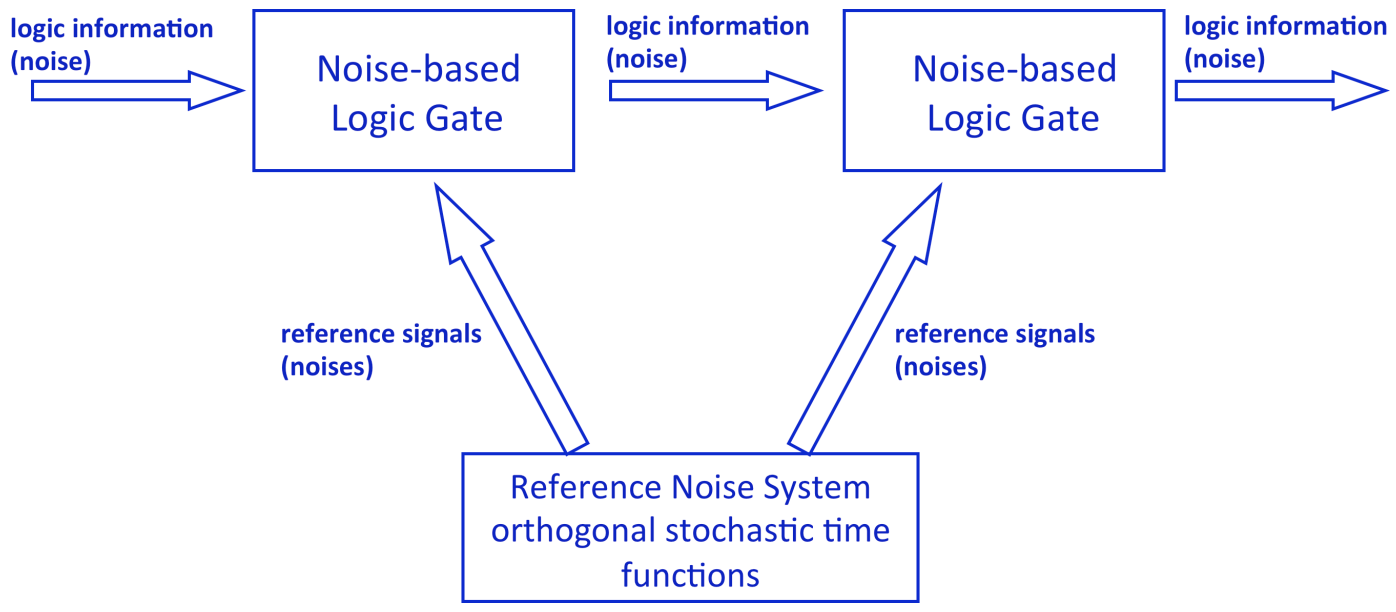

**Figure 1**. Generic noise-based logic hardware scheme.

The reference noises (in the infinite-time limit) can also be viewed as *orthogonal vectors;* thus superpositions of these signals represent a 2*M* dimensional geometrical *space*. Nonlinear operations on the reference noises and their *superpositions* (see Fig. 2) lead out from this original space and form a *hyperspace* with exponentially higher dimensions.

## *1.2. Hyperspace (Hilbert space) vectors: product-strings*

It is an important feature that the products of these noises form a *hyperspace* with $2^M$ geometrical dimensions that we describe below. This hyperspace forms the same Hilbert space that full-scale quantum computing is utilizing.

Numbers in the NBL system are *M*-bit binary numbers represented by product-strings formed by the noises of the corresponding bit values. For example, in the case of *M*=4, the signal of the binary number 1010 is carried by the noise product $R_{11}(t)R_{20}(t)R_{31}(t)R_{40}(t)$ .

Note that the above string corresponds to the vector |1010> in the quantum Hilbert space. This 4 noise-bit system can represent $2^4$ different binary numbers similarly to classical logic circuitry of 4 bits bit resolution. However, the major difference is that, similarly to full-scale quantum computers, the INBL system can carry simultaneously all these numbers in a parallel fashion, with a single signal which is the superposition of all these product-strings. INBL operations on these superpositions act simultaneously on each element in this superposition. This situation can lead to significant speedup.

### 1.2.1 Exponentially large superpositions with polynomial complexity

Many exponentially large superpositions can be treated by polynomial complexity in INBL, which is also a core feature of full-scale quantum computing. For example, the Universe *U*(*t*), which is the superposition of all the binary numbers, can be created by adding the signals of the Low (0) and the High (1) bit values of each noise-bit and then multiplying these sums:

$$U(t)=\left[R_{10}(t)+R_{11}(t)\right]\left[R_{20}(t)+R_{21}(t)\right]\ldots\left[R_{M0}(t)+R_{M1}(t)\right] \quad . \tag{2}$$

Even though this quantity can be set up by *M* addition and *M*-1 multiplication, that is, by 2*M*-1 elementary algebraic operations, when Equation 2 is expanded it forms the sum of all the 2*M* different product-strings (all the binary numbers) this *M* noise-bit system can form. For practical solutions for reducing the probability of zero amplitudes of the Universe, see Section 1.3.

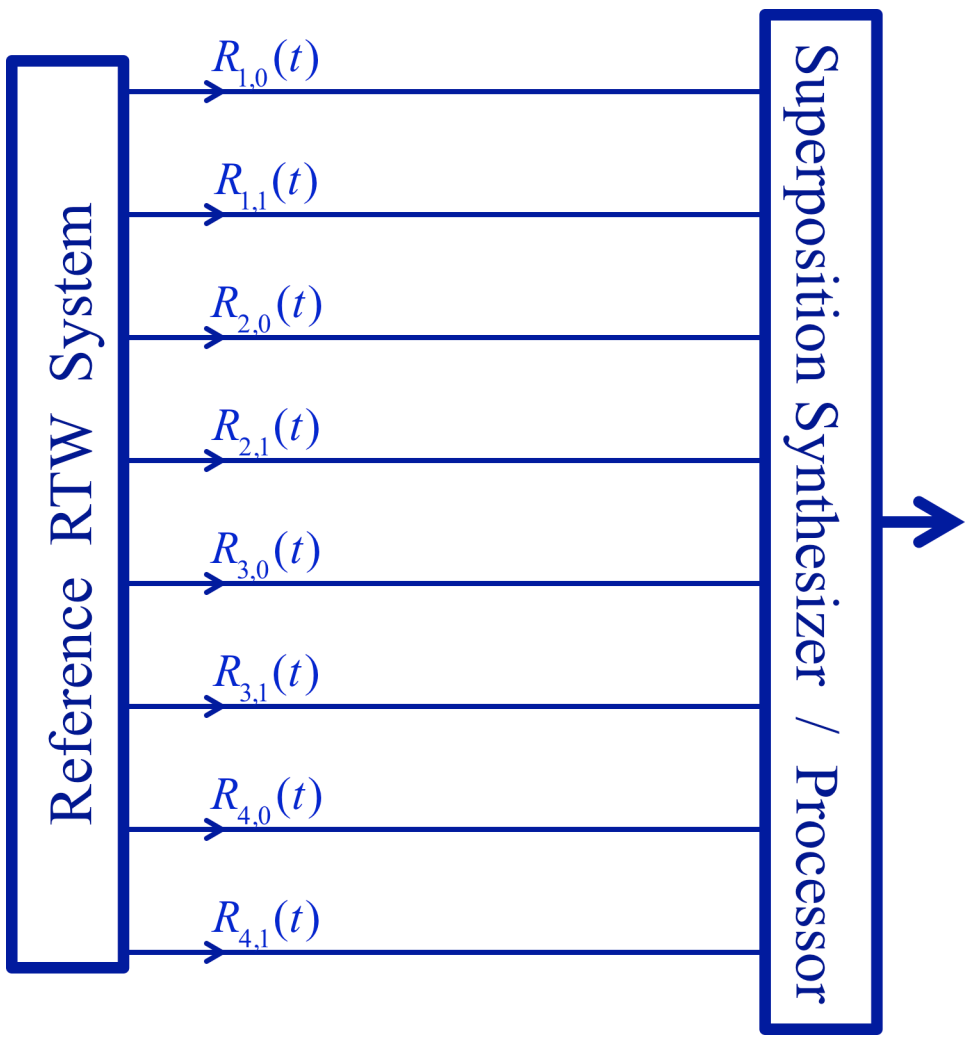

**Figure 2**. Generic hardware schematics to form a superposition from the reference signals.

It is important to note that although the same technique can be applied to deterministic signals, such as with sinusoidal waves as reference signals, such exponentially large superpositions will require exponential time complexity due to the frequency byproducts of multiplication, which results in an exponential growth of the bandwidth. Even though the bandwidth with continuum noises would increase due to the multiplication, linear frequency filters could limit the effect without losing the distinguishability of logic states; thus the polynomial complexity can be kept by using stochastic reference signals. More on this issue is in Section 1.3.

Another simple example is the superposition $Y_{\text{even}}(t)$ of all the even numbers, which can be obtained by skipping the term corresponding to the 1 bit value of the lowest bit:

$$Y_{\text{even}}(t) = R_{10}(t)\left[R_{20}(t)+R_{21}(t)\right] \dots \left[R_{M0}(t)+R_{M1}(t)\right] \quad , \tag{3}$$

and the superposition $Y_{\text{odd}}(t)$ of all the odd numbers is obviously just

$$Y_{\text{odd}}(t) = U(t) - Y_{\text{even}}(t) \quad . \tag{4}$$

Equation 4 represents an operation with an exponentially large, O($2^M$), input and an exponentially large, O($2^M$), output that does not overlap with the input. This requires requires only O($M$) complexity (see in Section 1.3) and it is carried out by a single algebraic operation: subtraction.

Note, full-scale quantum computers provide similar features in their Hilbert space with their unitary operators. The major difference between the quantum and the NBL schemes stems from the accessibility and robustness of the classical physical information, which not only offers but also requires different approaches in the algorithms. Thus we cannot say that noise-based logic and its algorithms are quantum-inspired. Yet we can say that many of the problems that NBL can address are quantum-inspired and both the quantum and the NBL schemes have the same limitations: small input and small output.

*1.3 Random telegraph waves as a Reference System*

Theoretically speaking, reference signals can be arbitrary types of stochastic processes, provided they are statistically independent and have zero mean.

For practical purposes, simple two-state binary noises – called Random Telegraph Waves (RTW) – have been proposed as reference signals, including standard ones with +1/-1 values; RTWs with different amplitudes, such as, +1/-1 for the High (1) bit values and +0.5/-0.5 for the Low (0) bit values (see Figs. 3-6) to avoid dominantly zeros in the amplitude of the Universe; or with complex values [10]; and time-shifted [9] versions that can additionally combine all the above ones. When the reference signals are independent RTWs, the products strings will also be RTWs with zero cross-correlation between different strings, see Fig. 5. These schemes have the advantage that binary multiplications and additions can be done accurately and economically by a digital computer while the bandwidth will remain the same order. Moreover, even though the amplitude values in a superposition can fluctuate over an exponentially large range, due to the binary handling of data, this exponential feature requires only a polynomial number of bits of resolution.

For example, in an RTW-based, $M$ noise-bit INBL system with amplitudes +1/-1 for the High (1) bit values and +0.5/-0.5 for the Low (0) bit values, the $R_{10}(t)R_{20}(t)...R_{M0}(t)$ string (corresponding to the ∣00...0> quantum state vector) has amplitude $2^{-M}$ and the amplitude of the Universe is bounded by $+2^{M}/-2^{M}$, see Figs. 3-6. For the schemes in the present paper, without giving up generality, we visualize this RTW system. A digital computer with 2$M$-bit resolution, which easily can be satisfied with proper software, can accurately handle this system. For $M$ in the range of thousands implies some slowdown; however, that slowdown is only polynomial in $M$.

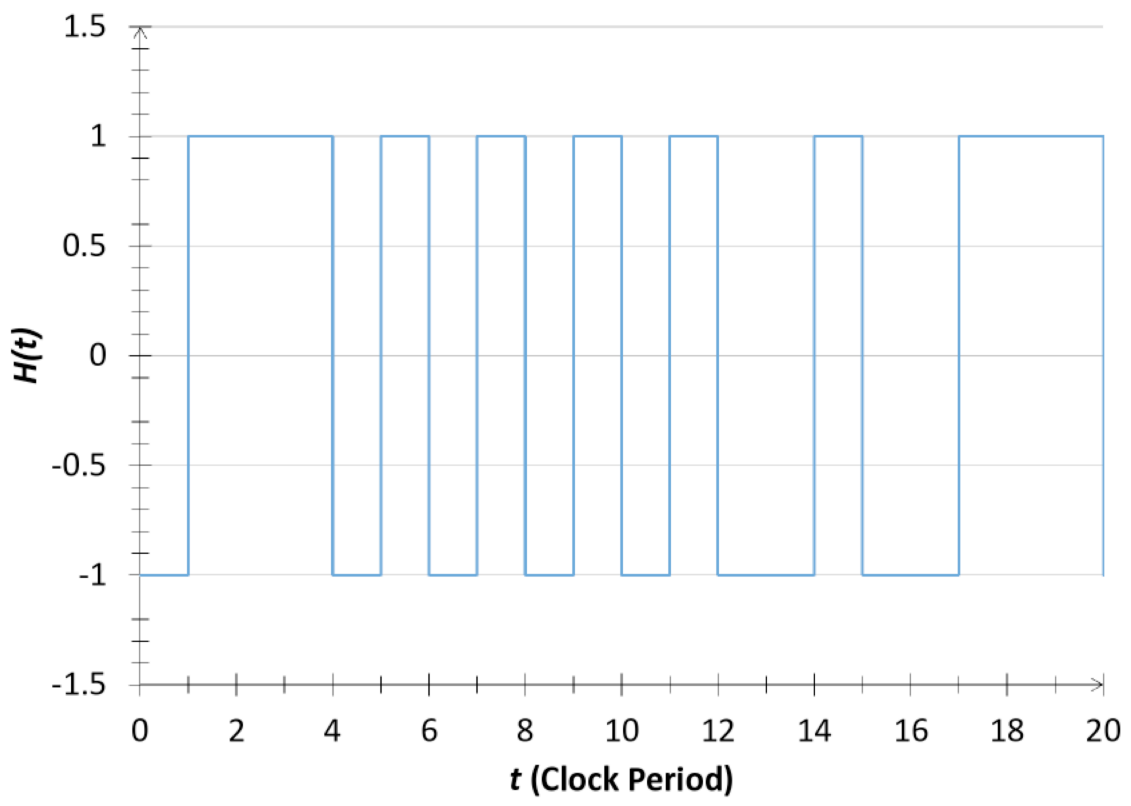

**Figure 3**. Computer simulation [14] of a random telegraph wave carrying the High-bit value of a noise-bit in the asymmetric INBL scheme.

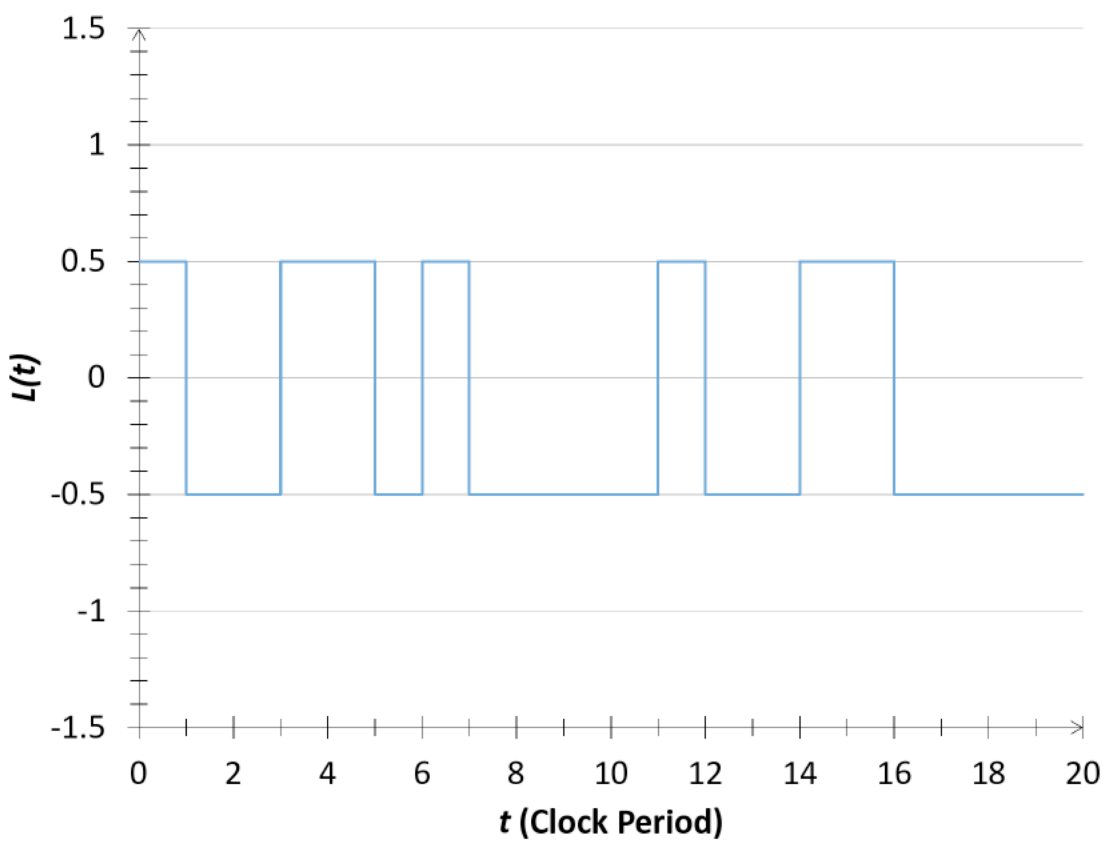

**Figure 4**. Computer simulation [14] of a random telegraph wave carrying the Low-bit value of a noise-bit in the asymmetric INBL scheme.

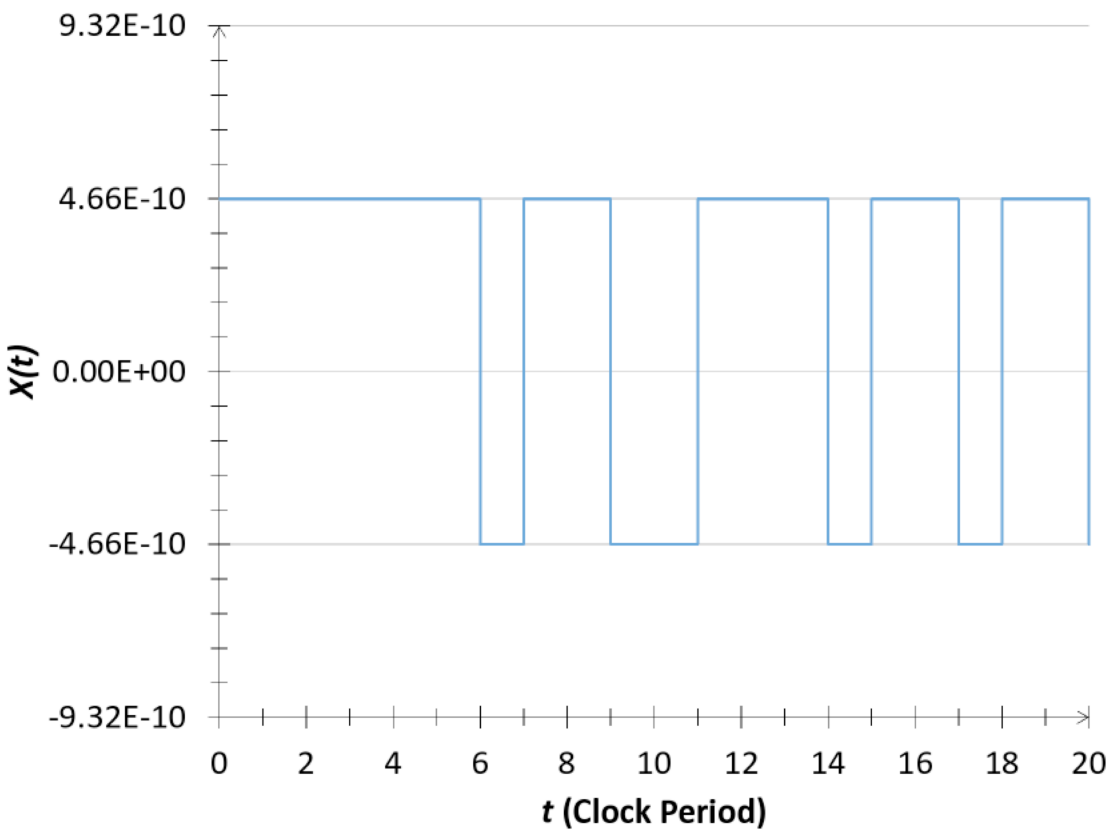

**Figure 5**. Computer simulation [14] of the product-string of the number 2 (binary number 00000000000000000000000000000010) in an $M = 32$ noise-bit INBL scheme.

In conclusion, an INBL processor is envisioned as a classical PC computer with proper software to handle $2M$-bit resolution and the four basic arithmetic calculations, and a physical true random number generator (TRNG).

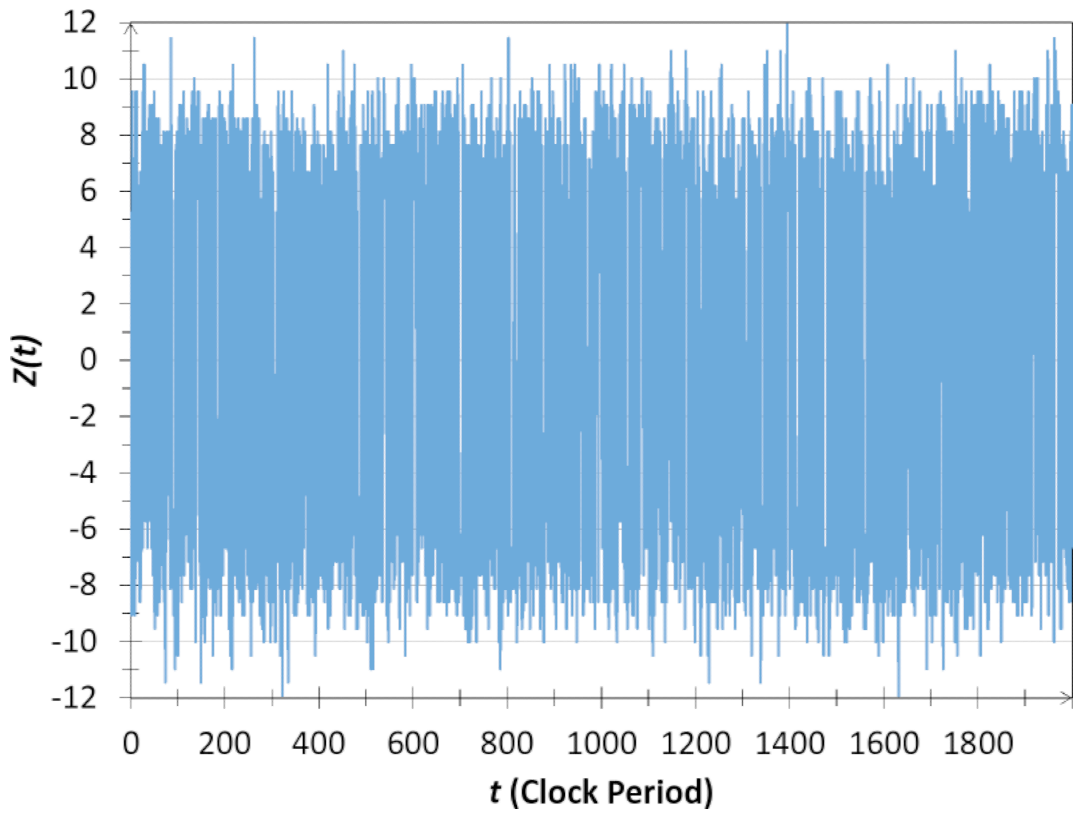

**Figure 6**. Computer simulation of the $N = 32$ noise-bit universe in the asymmetric INBL scheme over 2000 clock periods. The time function $U(t)$ of the universe, which represents the superposition of all integer numbers from 0 to 4294967295, is plotted in a logarithmically distorted way, defined by $Z(t) = \mathrm{sign}\left[U(t)\right] \log\left[2^{32}\left|U(t)\right|\right]$, for better visibility of small and large variations.

Note: For a classical physical TRNG, the thermal noise of resistors offers the best choice due to the quantum randomness of the phonon scattering of electrons and holes and the extraordinarily large degrees of freedom of these scattered charge carriers resulting in a classical physical enhancement of quantum randomness. Known quantum random generators lack the second feature and they are prone to bias due to various fluctuations in their control degree of freedoms, such as polarizer/beamsplitter angle vibrations, polarization noise of lasers, 1/f noise, etc.. Thus classical physics offers advantageous potential for TRNG yet their design still requires special attention and possibly hashing.

## 2. New measurement tools involving the reference wires

It has been recognized earlier in the context of "ghost states" [10] and CNOT gates [15] acting on superpositions that the manipulation of the reference system is a powerful computational tool. The new results in this paper also utilize this technique in a different fashion. We suppose that in the $M$-bit INBL system the unknown and unsorted superposition consists of an arbitrary sum of distinguishable $M$-noise-bit product-strings (see Sections 1.2 and 1.3).

*2.1 "Quantum" measurement of a given string within a superposition: "collapse of the system wavefunction"*

Any of the $M$-long strings $X(1), X(2)...X(M)$ in an $M$-noise-bit INBL system is using half of the reference wires to form its signal:

$$R_{1X(1)}(t)R_{2X(2)}(t)...R_{MX(M)}(t) \quad . \tag{5}$$

We can reduce an arbitrary superposition to any of its product-string elements by "grounding" (forcing a zero signal amplitude onto) the inverse (that is, other half) of the reference wires, namely, the reference wires of the bit values $\bar{X}(1), \bar{X}(2)...\bar{X}(M)$, while keeping the rest of the reference wires, the $X(1), X(2)...X(M)$ wires, at their reference sources. For the required hardware, see Figure 7, and for a practical example, see Figure 8. Any other string in the arbitrary superposition uses at least one of the grounded bit values thus the signal components all these product-strings become zero in the superposition. This situation corresponds to the "collapse of the system wavefunction" in a quantum system. If the superposition signal amplitude is not zero during the clock period when the measurement is carried out, and after the grounding the signal becomes zero, that is a proof that the product-string was not in the superposition. Otherwise the new signal will be a non-zero signal $R_{1X(1)}(t)R_{2X(2)}(t)...R_{MX(M)}(t)$ of the product-string.

Therefore, grounding the inverse reference wires of a string and observing the remaining signal of the superposition represents the *quantum-type measurement* in instantaneous noise-based logic. Note, there are other types of measurements, too, but they have statistical nature while grounding the reference wires yields a *deterministic measurement* provided the instantaneous amplitude of the original superposition is non-zero at that time, that is, during the clock-period of the measurement.

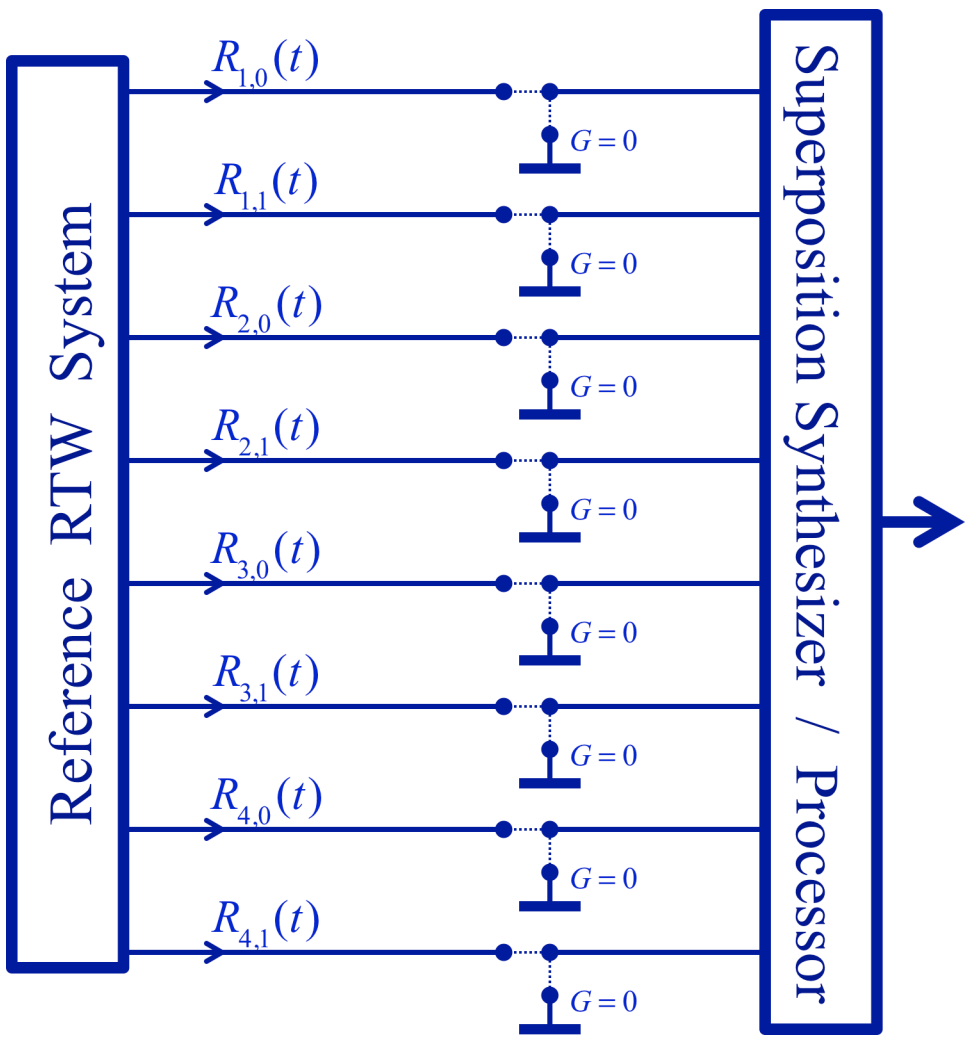

**Figure 7**. INBL generic hardware scheme with two-state switches in the reference wires to provide the opportunity of grounding (that is, forcing zero onto) selected reference wires.

It is important to note that, even though this measurement looks similar to quantum measurements, it is *very different from quantum measurements*. *Quantum measurements are statistical* which means that they not only *do contain errors* but they also *do require an exponential complexity* when the task is to measure a string in an exponentially large superposition. For example, in a photonic quantum system with $M$ qubits, the Universe consists of the superposition of $2^M$ different strings. If in this uniform (flat) superposition, we want to detect a single string with acceptable error probability, the number $P$ of available photons in the system must satisfy:

$$P >> M2^M \tag{6}$$

implying an *exponential time complexity in the measurement system of filters/beamsplitters and coincidence detectors with finite time-resolution*. On the other hand, in INBL the same measurement takes just a single time step.

*2.2 Deterministic measurement and evaluation of entangled states and their superposition in INBL*

Suppose Alice sets up a two-noise-bit system ($M$=2) with a superposition signal $S(t)$ containing entanglement (similar to the Einstein-Rosen-Podolsky [1] problem's superposition and the related Bell states):

$$S(t) = R_{10}(t)R_{21} + R_{11}(t)R_{20}(t) \quad . \tag{7}$$

This 0(1),1(2) + 1(1),0(2) superposition is unknown by Bob except the conditions that it consists of the sum of any of the possible $R_{ij}(t)R_{pq}(t)$ two-bit strings $i,p \in \{1,2\}; j,q \in \{0,1\}$, with the restriction that a given value of a given bit can exist at most in a single string. For example, alternative superpositions could be 0(1),0(2) + 1(1),1(2) , that is, $R_{10}(t)R_{20}(t) + R_{11}(t)R_{21}(t)$; or single strings such as 0(1),1(2) , that is, $R_{10}(t)R_{21}(t)$; 1(1),0(2) , that is, $R_{11}(t)R_{20}(t)$; 0(1),0(2) , that is, $R_{10}(t)R_{20}(t)$; or 1(1),1(2) , that is, $R_{11}(t)R_{21}(t)$.

Bob's job is to measure $S(t)$ and decide which is the actual superposition from these different possibilities. To identify strings statistical methods exist [e.g. 9,11,12,14] and some of them, like the time-shifted RTW scheme [9] and the Stacho [11] method require only polynomial computation complexity versus the number of noise-bits. However, these methods used on superposition would require *exponential complexity* with expanding $M$ (except when the job is to identify a single string and the rest of the superposition is known [14]).

Here we show that, by utilizing operations on the reference wires, the measurement and analysis can be done in a fast way, which is deterministic, with polynomial complexity. The procedure takes place during a clock time $t$ when the superposition signal amplitude is non-zero, that is, $S(t) \neq 0$ :

i) Bob first checks the existence of the reference signal $R_{10}(t)$ in the superposition $S(t) \neq 0$ by grounding (forcing zero on) the reference wire of the *inverse bit value* of $R_{10}(t)$, that is, forcing $R_{11}(t) = 0$, and then checking if the resulting superposition signal $S_2(t)$ is zero. In the above example, see Equation 7, the superposition gets reduced to:

$$S_2(t) = R_{10}(t)R_{21}(t) \quad , \tag{8}$$

which means $S_2(t) \neq 0$.

In alternative cases, when the grounding causes $S_2(t) = 0$ Bob jumps to step (iii) below. In the present situation, $S_2(t) \neq 0$ (see Equations 7 and 8), Bob needs to evaluate which value of the second bit is entangled with the 0 value of the first bit. This string can be evaluated in various known statistical ways [e.g. 9,11,14] but here we again propose a deterministic tool by utilizing the reference wires:

ii) While still being in the same clock period $t$ and keeping $R_{11}(t) = 0$, Bob grounds also the 0 value of the second bit, $R_{20}(t) = 0$, and gets the superposition $S_3(t)$. If $S_3(t) = 0$ then $R_{20}(t)$ is the entangled bit signal. If $S_3(t) \neq 0$ then $R_{21}(t)$ is the bit value in question, which is the case in the present example.

Note, Bob can proceed in an alternative way: he grounds the reference $R_{21}(t)$, $R_{21}(t)=0$, while keeps the grounding $R_{11}(t)=0$. If $S_3(t)=0$ then $R_{21}(t)$ represents the entangled bit value.

iii) Now, Bob repeats the above procedures (i-ii) with swapped bit values. He first checks the existence of the reference signal $R_{11}(t)$ in the superposition by grounding the reference wire of the inverse bit value (forcing $R_{10}(t)=0$). If the reference signal $R_{11}(t)$ is not in the original superposition (Equation 7) then $S_2(t)=0$ and Bob completed the evaluation. Otherwise he proceeds with the swapped bit values as described in the above procedure (i-ii).

*2.3 Essential differences between measuring entanglement in INBL and quantum systems*

The above measurement schemes can be expanded and generalized for combined superpositions that entangle different superpositions into a jointly entangled one; see the search algorithms presented in Section 3. The particular differences from measurement of quantum entanglement are the *deterministic* nature of the INBL schemes in a similar fashion as it is described above in Section 2.1.

In the case of the quantum measurement of the entangled bit-1 in the superposition described by Equation 7, the measurement would detect the 0 and 1 values of bit-1 with 0.5 probability. On the other hand, in INBL we choose which value of bit-1 we want to check, the 0 or the 1 value. This is a major difference because, in an exponentially large superposition (see Section 3), the quantum scheme would require an exponentially large number of measurements in the case of a uniform distribution such as the one in Equation 7. This is the reason why Grover's quantum search algorithm transforming the originally uniform distribution to a new superposition where the amplitude of the searched term is enhanced. In this way, a minor $N^{0.5}$ speedup is reached but that still keeps exponential time requirement on an exponential superposition. Section 3 shows measurements that offer exponential speedup with instantaneous noise-based logic.

## 3. The two search algorithms

The two search algorithms presented below are based on the application and generalization of the measurements described in sections 2.1 and 2.2, respectively.

*3.1. Search for the existence of a string or string-fragment in an unknown, unsorted superposition*

Here we again suppose that, in the $M$-bit INBL system, the unknown and unsorted superposition consists of an arbitrary sum of $M$-long product-strings. This method is the generalization of the scheme described in Section 2.1. It is exponentially, $\frac{2^M}{M}$, faster than known classical alternatives, and $\frac{2^M}{M^{1.5}}$ faster than Grover's for the problems in this section. We note that with a suitably complex oracle Grover's algorithm can be applied to other problems outside the scope of the method presented here.

*Method*:

The whole search process is carried out during a clock period $t$ when the original superposition is non-zero, $S(t)\neq 0$.

I) Ground the reference wires of the inverse bit values of the search term;

II) Check if the remaining superposition is non-zero. If it is nonzero, the search string or string fractions exist in the superposition.

*Example-1, full string search*: suppose that *M*=4 and the job is to find out if the 1(1),0(2),1(3),0(4) string (represented by the $R_{11}(t)R_{20}(t)R_{31}(t)R_{40}(t)$ product) is an element of the otherwise unknown superposition $S(t)$. Let us assume that the superposition is

$$S(t) = R_{11}(t)R_{20}(t)R_{31}(t)R_{40}(t) + R_{10}(t)R_{20}(t)R_{31}(t)R_{40}(t) + R_{10}(t)R_{21}(t)R_{31}(t)R_{40}(t) \quad . \tag{9}$$

The inverse of the 1(1),0(2),1(3),0(4) string is 0(1),1(2),0(3),1(4) ; therefore we ground the corresponding reference wires, see Figure 8:

$$R_{10}(t) = 0; \;\; R_{21}(t) = 0; \;\; R_{30}(t) = 0; \;\; R_{41}(t) = 0 \quad . \tag{10}$$

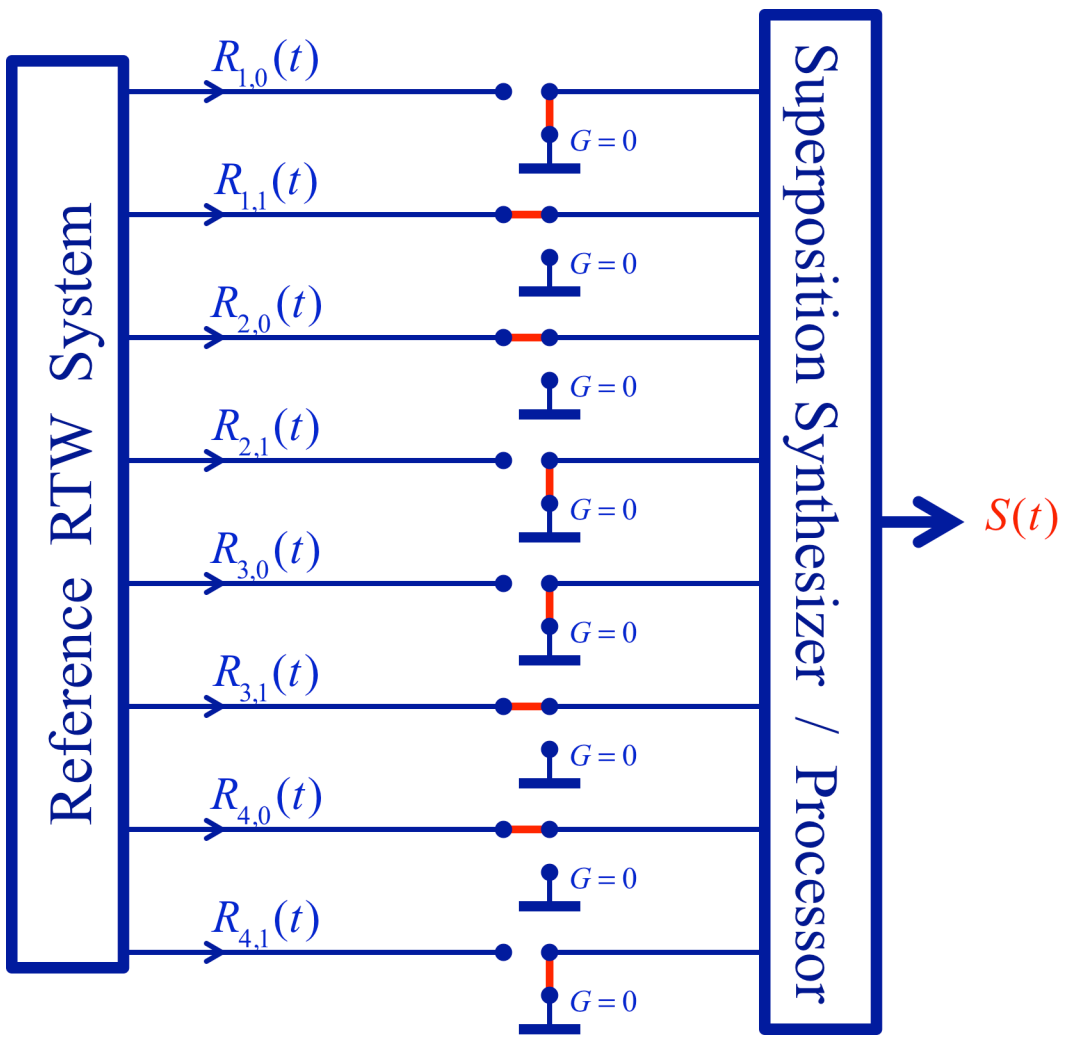

**Figure 8**. Switch arrangement for the instantaneous search of the 1(1),0(2),1(3),0(4) string in the superposition. The reference wires of the 1(1), 0(2), 1(3) and 0(4) bit values are kept at their reference sources while the inverse bit values are grounded. If that string is not in the superposition, the superposition signal output *S*(*t*) will become zero.

In any of the other possible *M*-long strings in the *M*-bit system, at least one of the reference noises listed in Eqs. 10 will be present. That means that only the search string is non-zero, and if the remaining superposition is non-zero, such as in this case, the search string is present in the superposition.

The operation requires *M*=4 switching thus the time complexity is O(*M*)=4. The maximum size of the superposition is $2^M$ = 16. The speedup is $2^M / M = 4$.

*Example-2, string fragment search with single string result*: Search for the 0(1),0(2), AND 0(4) string fragments in the superposition. The inverse bit values are 1(1),1(2),1(4) therefore we ground the corresponding reference wires:

$$R_{11}(t) = 0; \;\; R_{21}(t) = 0; \; R_{41}(t) = 0 \quad . \tag{11}$$

The superposition in Equation 9 will be non-zero (because of the second term) indicating the simultaneous existence of the searched string fragments within at least one of the strings.

*Example-3, string fragment search with superposition result*: Search for the same, $0(1),0(2),$ AND $0(4)$, string fragments in this modified superposition:

$$S'(t) = R_{11}(t)R_{20}(t)R_{31}(t)R_{40}(t) + R_{10}(t)R_{20}(t)R_{31}(t)R_{40}(t) + \\ + R_{10}(t)R_{20}(t)R_{30}(t)R_{40}(t) + R_{10}(t)R_{21}(t)R_{31}(t)R_{40}(t) \quad . \tag{12}$$

We do the same reference wire grounding as in Eq. 11 above, this time of bit values $1(1),1(2),1(4)$. In this case both the second and the third terms in Equation 12 will be non-zero and the remaining superposition will be:

$$S_2^{'}(t) = R_{10}(t)R_{20}(t)R_{31}(t)R_{40}(t) + R_{10}(t)R_{20}(t)R_{30}(t)R_{40}(t) \quad . \tag{13}$$

In similar cases, depending of the type of reference noises, there may be clock periods when the absolute values of the remaining product strings form $S_2^{'}(t) = 0$. In such cases this method becomes statistical. With (asymmetric) random-telegraph waves the worst-case error probability scales as $\varepsilon \propto 2^{-\tau}$ where $\tau$ is the number of observational clock steps, because the probability that two independent RTW's go with the same absolute value, and opposite signs, over $\tau$ steps is $2^{-\tau}$ (see string verification in [12]). As soon as the $S_2^{'}(t)$ signal becomes non-zero, it is proven that the searched string fragments are present and the error probability of this decision is zero.

### *3.2 Phonebook search*

Suppose we have phonebook with *n* names of *N*-bit resolution and *s* different phone numbers with *S*-bit resolution. If inverse lookup is not needed, the same phone number may belong to several names, thus $s \le n$. Note, $n = 2^N$ but *S* is arbitrary with the obvious restriction $S \ge \log_2 s$.

In this system, there are two related and independent INBL reference systems with *N* and *S* noise-bits, respectively, that is, 2*N*+2*S* independent noise sources with zero mean. The product strings of the name database are entangled with the product strings of the phone number database. The phonebook superposition is then:

$$S_{\mathrm{PB}}(t) = \sum_1^n N_i(t)S_i(t) = \sum_1^n B_i(t) \quad , \tag{14}$$

where $N_i(t)$ is the *i*-th name string and $S_i(t)$ is the corresponding entangled phone number component; $N_i(t) \ne N_j(t)$ for $i \ne j$ but $S_i(t)$ and $S_i(t)$ maybe identical for $i \ne j$ (when inverse lookup is not needed). The phonebook string $B_i(t)$ is the product of the corresponding name and phone number strings:

$$B_i(t) = N_i(t)S_i(t) \quad . \tag{15}$$

After these preparations, the phonebook search algorithm is obvious and it is equivalent with Example-2, the string fragment search with single string result, shown in Section 3.1.

- Search for the existence of the string-fragment that is identical to $N_i(t)$, that is, ground the reference wires of the inverse bit values of $N_i(t)$. Then the $S_{\mathrm{PB}}(t)$ original superposition wave collapses into the single string, the $B_i(t) = N_i(t)S_i(t)$ product. This operation requires $N$ switching operations ($N$ groundings).

- Then, while keeping those $N$ references wires grounded, ground *one by one* the $2S$ reference wires of the phone number database and record which bit value groundings result in $B_i(t) = 0$. These bit values constitute the phone number string $S_i(t)$.

The time and hardware complexity of this search algorithm is $O\left[\log_2 n + 2S\right]$ in general, where $n$ is the size of the database and $S$ is the bit resolution of the phone numbers and it is $O\left[3\log_2 n\right]$ when $S = \log_2 s$.

In the case of one-to-one correspondence between the names and phone numbers ($n$=$s$), the algorithm obviously offers inverse phonebook search, too. To do that search the name and phone number strings in the above description must be swapped.

## 4. Discussion and Conclusions

Classical stochasticity via the instantaneous noise-based logic schemes may offer more than quantum computers because of the lack of the fundamental disadvantages of quantum systems [19] which are the decoherence, the statistical nature of quantum measurements, and the fact that measurements destroy the state. Examples of this superiority are shown in the present paper where, once the search has begun, the results come out deterministically without statistical errors, resulting in exponential speedup compared to Grover's quantum search algorithm.

To begin the search, a single clock period with non-zero superposition amplitude is required. The whole search is done during this single period. Thus a new research problem has emerged: to find optimal reference signal systems where the superposition signals are zero with very low probability and, if they take the zero value, this value quickly changes to non-zero during the subsequent clock periods. Tested superposition amplitudes spend very short times at zero, see Figure 6, however a careful study of this issue is warranted. Zero superposition amplitude may introduce a random waiting time (typically with exponentially decaying probability to remain at zero, see [12]) before the search can begin; however, that does not change the fact that the search result is error-free and deterministic.

The present paper is another demonstration that, despite the resemblance between instantaneous noise-based logic and quantum computing, the NBL schemes and their modes of operations are substantially different even if the achieved goals can be the same. While explorations of NBL are still taking place, it is already clear that the robustness and controllability of classical physical systems and information may sometimes offer competitive advantages over quantum informatics. The results in this paper show examples where NBL schemes not only outperform a known corresponding quantum algorithm but that can also achieve the goal with fewer resources. It is important to emphasize that the schemes shown in this paper *are not quantum-inspired* and they utilize classical physical stochasticity.

**Author Contributions:** Conceptualization and methodology, L.B.K.; validation, L.B.K., W.D.; formal analysis, L.B.K.; investigation, L.B.K., W.D.; resources, L.B.K.; data curation, L.B.K.; writing—original draft preparation, L.B.K.; writing—review and editing, L.B.K., W.D.; visualization, L.B.K.; supervision, L.B.K., W.D.; project administration, L.B.K.; funding acquisition, L.B.K.

**Funding:** This research received internal Texas A&M funding in the form of a T3 grant (2019-2020).

**Acknowledgments:** We are grateful for the Reviewer for pointing out the potential problems of zero instantaneous amplitude of a superposition that needed to be addressed. Discussions with Michel Dyakonov, Andreas Klappenecker, Tamas Horvath and Suhail Zubairy are appreciated.

**Conflicts of Interest:** The authors declare no conflict of interest.